\documentclass[twocolumn,showpacs,prl,superscriptaddress,aps,amsmath,amssymb]{revtex4}
\usepackage{graphicx}

\begin{document}

\title{Interplay between magnetism and charge instabilities in layered NbSe$_{2}$}

\author{Simon Divilov}
\affiliation{Departamento de F\'{i}sica de la Materia Condensada,  Universidad Aut\'{o}noma de Madrid, Cantoblanco, 28049 Madrid, Spain.}
\affiliation{Condensed Matter Physics Center (IFIMAC), Universidad Aut\'{o}noma de Madrid, Cantoblanco, 28049 Madrid, Spain.}

\author{Wen Wan}
\affiliation{Donostia International Physics Center (DIPC), Paseo Manuel de Lardiz\'{a}bal 4, 20018 San Sebasti\'{a}n, Spain.}

\author{Paul Dreher}
\affiliation{Donostia International Physics Center (DIPC), Paseo Manuel de Lardiz\'{a}bal 4, 20018 San Sebasti\'{a}n, Spain.}

\author{Miguel M. Ugeda}
\email{mmugeda@dipc.org}
\affiliation{Donostia International Physics Center (DIPC), Paseo Manuel de Lardiz\'{a}bal 4, 20018 San Sebasti\'{a}n, Spain.}
\affiliation{Centro de F\'{i}sica de Materiales (CSIC-UPV-EHU), Paseo Manuel de Lardiz\'{a}bal 5, 20018 San Sebasti\'{a}n, Spain.}
\affiliation{Ikerbasque, Basque Foundation for Science, 48013 Bilbao, Spain.}

\author{F\'{e}lix Yndur\'{a}in}
\email{felix.yndurain@uam.es}
\affiliation{Departamento de F\'{i}sica de la Materia Condensada,  Universidad Aut\'{o}noma de Madrid, Cantoblanco, 28049 Madrid, Spain.}
\affiliation{Condensed Matter Physics Center (IFIMAC), Universidad Aut\'{o}noma de Madrid, Cantoblanco, 28049 Madrid, Spain.}

\date{\today}

\begin{abstract}
Using \textit{ab initio} methods based on density functional theory, the electronic and magnetic structure of layered hexagonal NbSe$_{2}$ is studied. In the case of single-layer NbSe$_{2}$ it is found that, for all the functionals considered, the magnetic solution is lower in energy than the non-magnetic solution. The magnetic ground-state is ferrimagnetic with a magnetic moment of 1.09 $\mu_{B}$ at the Nb atoms and a magnetic moment of 0.05 $\mu_{B}$, in the opposite direction, at the Se atoms. Our calculations show that single-layer NbSe$_{2}$ does not display a charge density wave instability unless a graphene layer is considered as a substrate. Then, two kinds of 3$\times$3 charge density waves are found, which are observed in our STM experiments. This suggest that the driving force of charge instabilities in NbSe$_{2}$ differ in bulk and in the single-layer limit. Our work sets magnetism into play in this highly-correlated 2D material, which is crucial to understand the formation mechanisms of 2D superconductivity and charge density wave order. 
\end{abstract}


\maketitle

Metallic layered transition metal dichalcogenides (TMD) have been studied since the early work of Wilson et al. \cite{Wilson-1,Wilson-2}  in the mid-seventies when it was discovered that, due to their quasi two-dimensional (2D) character, they present charge density waves (CDW) instabilities in addition to a superconducting ground state at moderate low temperatures. In the prototypical 2H-NbSe$_{2}$, the appearance of a 3x3 CDW has been widely studied both theoretically and experimentally \cite{Weber-PRB, Weber-PRL}, and has been attributed either to nesting of the Fermi surface or to a strong electron-phonon interaction (see for instance  \cite{Mazin-1, Mazin-2}). However, the direct mechanism has not been unambiguously established so far. The mechanism responsible for the superconducting ground state has been, however, solely attributed to the electron-phonon coupling. Recent experiments have made possible to explore the ultimate single-layer limit of NbSe$_{2}$ to reveal that both the 3x3 CDW and superconductivity persist, although the latter at lower temperatures with respect to the bulk \cite{Ugeda-NP}. It has been generally assumed that the mechanism of both CDW and superconductivity in bulk and single-layer NbSe$_{2}$ should be similar \cite{Calandra-few-layer, Ordejon-Ugeda}.

In addition to these collective electronic phases, it has been recently suggested that several TMD materials host magnetic order, thus enriching even more their phase diagrams. Early theoretical works based on first-principles calculations first predicted a ferromagnetic ground state of vanadium-based TMDs -  VX$_{2}$ (X = S, Se, Te)  - and, more recently, in other TMDs - MnSe$_{2}$ and CrSe$_{2}$(Te$_{2}$) - in the single-layer form\cite{Magnetism-TM-Se2, Ma, Attaca}. Significant but yet scarce experimental progress has already been achieved here \cite{Batzill, Fumega, Ohara, Avsar}. Among the TMD materials, the plausible existence of long-range magnetic order is particularly relevant in NbSe$_{2}$ since it could coexist with 2D superconductivity, which would dramatically impact the nature and fundamental properties of the superconducting state. Spin-integrated calculations reported antiferromagnetic order in single-layer NbSe$_{2}$ provided the CDW is absent, as a result of a competition between these two phenomena \cite{Strain, Zheng}. These calculations, however, lack of spin-resolved information at the DFT level and, therefore, a more exhaustive theoretical analysis using different levels of theory becomes crucial to accurately explore these effects. 

\begin{figure}[t]
\includegraphics[width=85mm]
{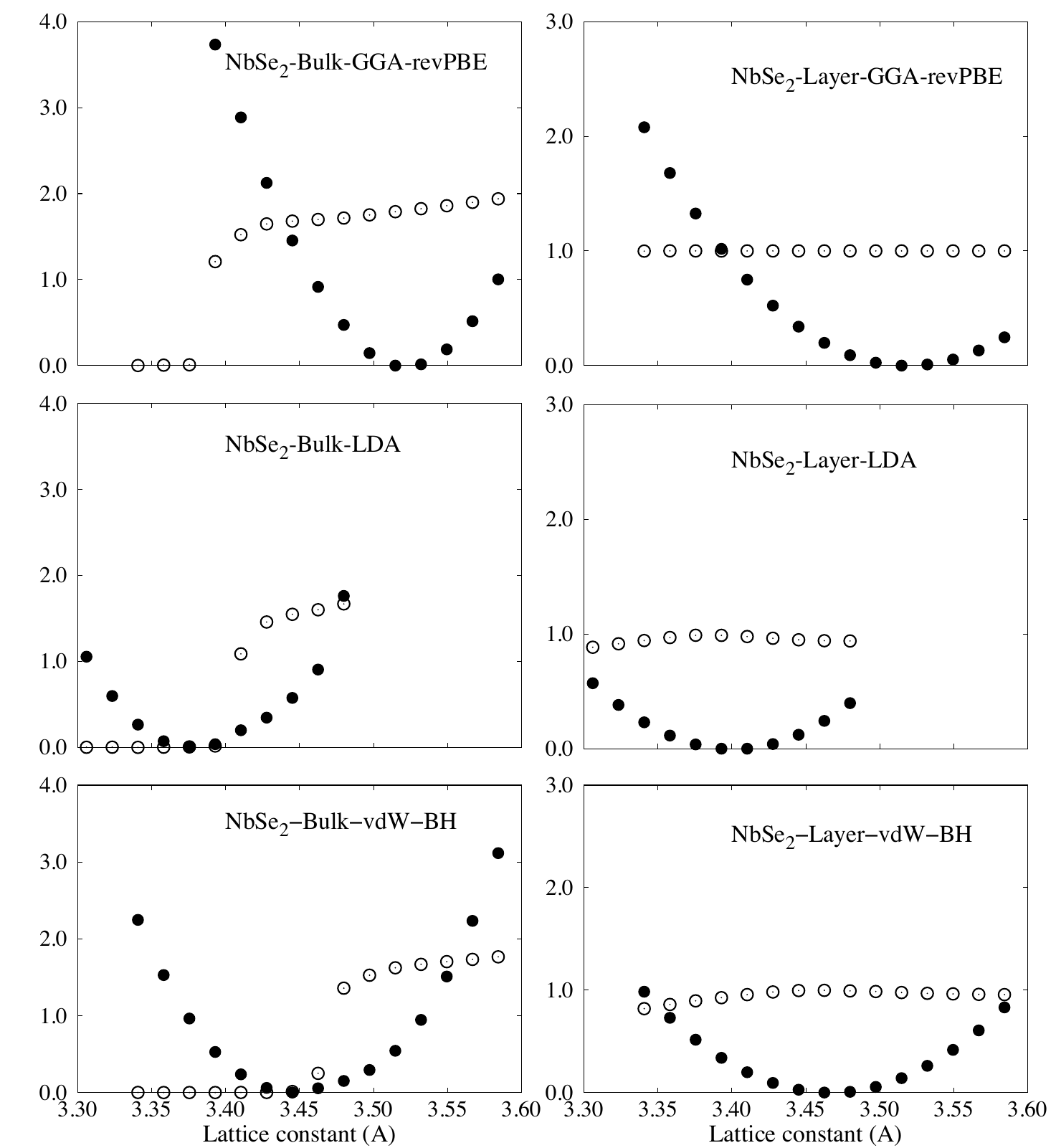}
\caption{Total energies (solid circles), in eV$\times$10, and total magnetic moment per unit cell (open circles), in Bohr magnetons $\mu_{B}$, versus lattice constant calculated for both bulk (left) and layered (right) 2H-NbSe$_{2}$. The calculations are done for the three different functionals as indicated.}
\label{Energies}
\end{figure}

Here we report combined density functional calculations and scanning tunneling microscopy and spectroscopy (STM/STS) measurements to reveal the geometrical and electronic structure of NbSe$_{2}$. The density functional theory (DFT) calculations \cite {DFT1, DFT2} were performed using the SIESTA code \cite {Siesta1, Siesta2} which uses localized orbitals as basis functions  \cite{Orbitals}. The calculations were done using a polarized double $\zeta$ basis set, non-local norm conserving pseudopotentials and, for the exchange correlation functional, in most of the calculations, we used the generalized gradient approximation (GGA)\cite{GGA} including the van der Waals interaction with the functional developed by Berland and Hyldgaard (BH) \cite {BH} and implemented by Rom\'{a}n-P\'{e}rez and Soler \cite {Roman}.

The calculations were performed with a stringent convergence criteria. The density matrix tolerance of the self-consistency, 2D Brillouin zone sampling, energy cut-off and force tolerance are $10^{-5}$, 3000 $k$-points, 1000 Ry and $1\times10^{-3}$ eV/\AA, respectively. Due to the rapid variation of the density of states (DOS) at the Fermi level, we used a polynomial smearing method \cite{smearing}. The calculations were done allowing for different spin-up and spin-down populations. Most of the calculations were done assuming possible collinear magnetic arrangements. In some instances, the possibility of non-collinear arrangements was considered, however no stable magnetic ground state was found. The low-temperature STM/STS experiments were carried out in a commercial (Unisoku) low-temperature UHV-STM system. Single-layer NbSe$_{2}$ samples were previously grown by molecular beam epitaxy in a UHV system (see SM\cite{SM} for detailed information regarding the sample growth).

We show the results of the total energy and magnetic moment of both bulk and single-layer hexagonal NbSe$_{2}$ for different functionals in Fig. \ref{Energies} . We first notice that for the commonly used GGA revPBE \cite{GGA} functional the bulk is magnetic at the equilibrium lattice constant. However, in the local density approximation (LDA) \cite{LDA2} and the van der Waals functional of Berland and Hyldgaard \cite {BH} with a modified treatment of the exchange, the bulk is non-magnetic although close to the magnetic solution. The lattice constant with the BH functional is 3.440 \AA{}  that compares well with the experimental one of 3.442 \AA \cite{Meerschaut}. Most of the subsequent calculations are done using this functional. In all the calculations spin-orbit coupling is not included since its effect is believed to be small \cite{Spin-Orbit}.

The situation for the monolayer is different. For all the functionals used, the system is magnetic at the equilibrium lattice constant and for a wide range of lattice constants, the magnetic moment being essentially independent of the lattice constant. The magnetic moment at the Nb atoms is around 1.09 $\mu_{B}$ whereas the magnetic moment at the Se atoms, pointing in the opposite direction, is around 0.05  $\mu_{B}$. This magnetic configuration is similar to the family of transition metal dichalcogenides \cite{Magnetism-TM-Se2}. The possibility of magnetism in layered NbSe$_{2}$ has not been considered before, except in the case of layers under strain \cite{Strain}.

\begin{figure}[h]
\includegraphics[width=85mm]
{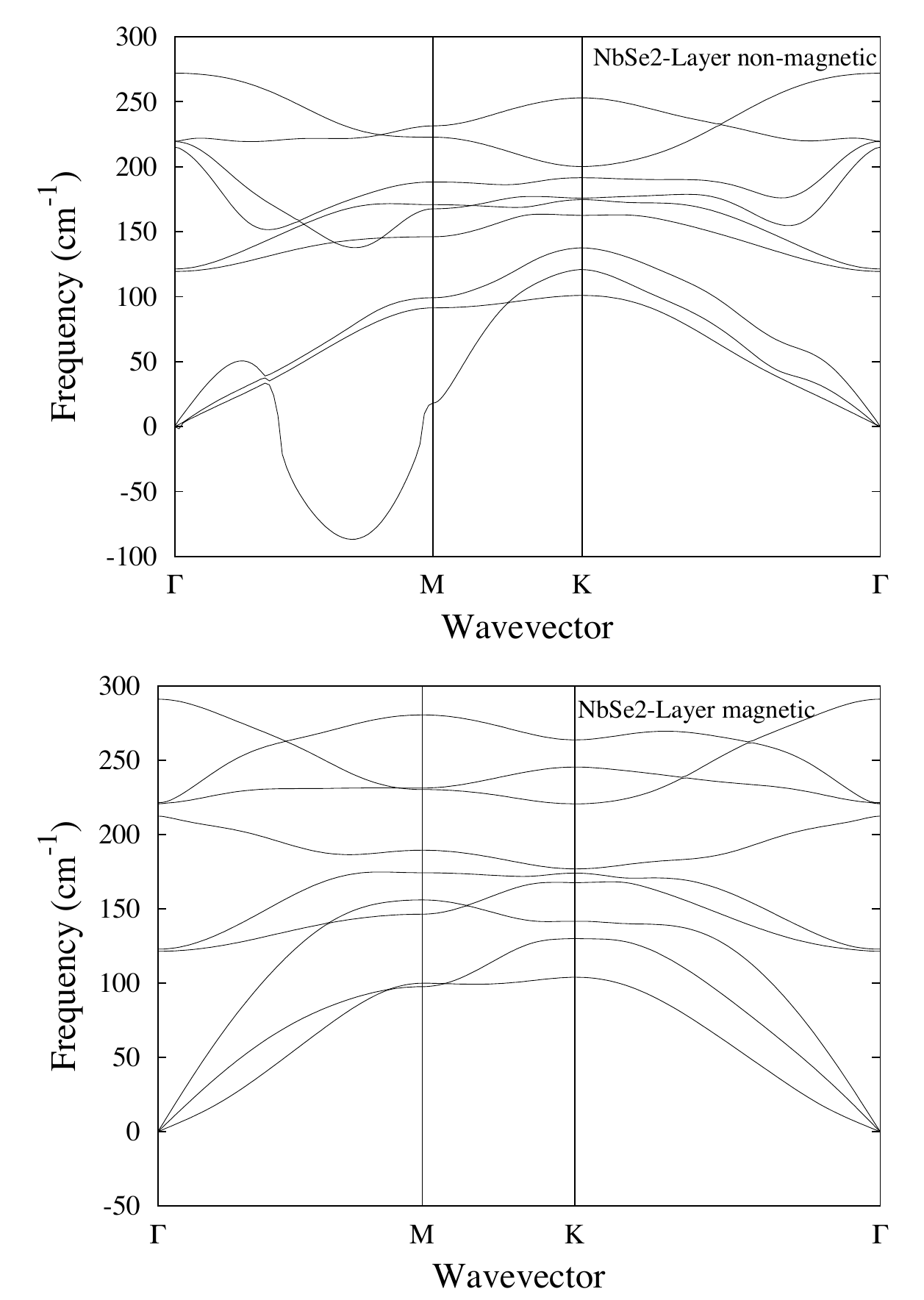}
\caption{Phonon dispersion relation of single-layer NbSe$_{2}$ calculated with the BH \cite{BH} functional. Top and bottom panel show the bands calculated for the non-magnetic and magnetic solution, respectively.}
\label{Phonons}
\end{figure}

To study the possible lattice instabilities we have calculated the phonon dispersion of single-layer NbSe$_{2}$, shown in Fig. \ref{Phonons}. As in the previous work \cite{Calandra-few-layer, Ordejon-Guinea, Ordejon-Ugeda,ion}, we find that for the non-magnetic calculation, the phonon spectrum shows a clear imaginary frequency (negative in the figure) compatible with a 3$\times$3 structural instability. However, for magnetic calculations, there were no soft phonon modes, indicating a stable structure with no structural distortions. We have also done a calculation with a 3$\times$3 supercell to look for any charge and/or lattice instabilities. Indeed, we find that the non-magnetic calculation exhibits atomic distortions whereas the magnetic solution is \textit{flat and undistorted with no sign of any charge instability}. The total energy is found to be 160 meV (per f.u.) lower than the distorted non-magnetic solution.

\begin{figure}[h]
\includegraphics[width=85mm]
{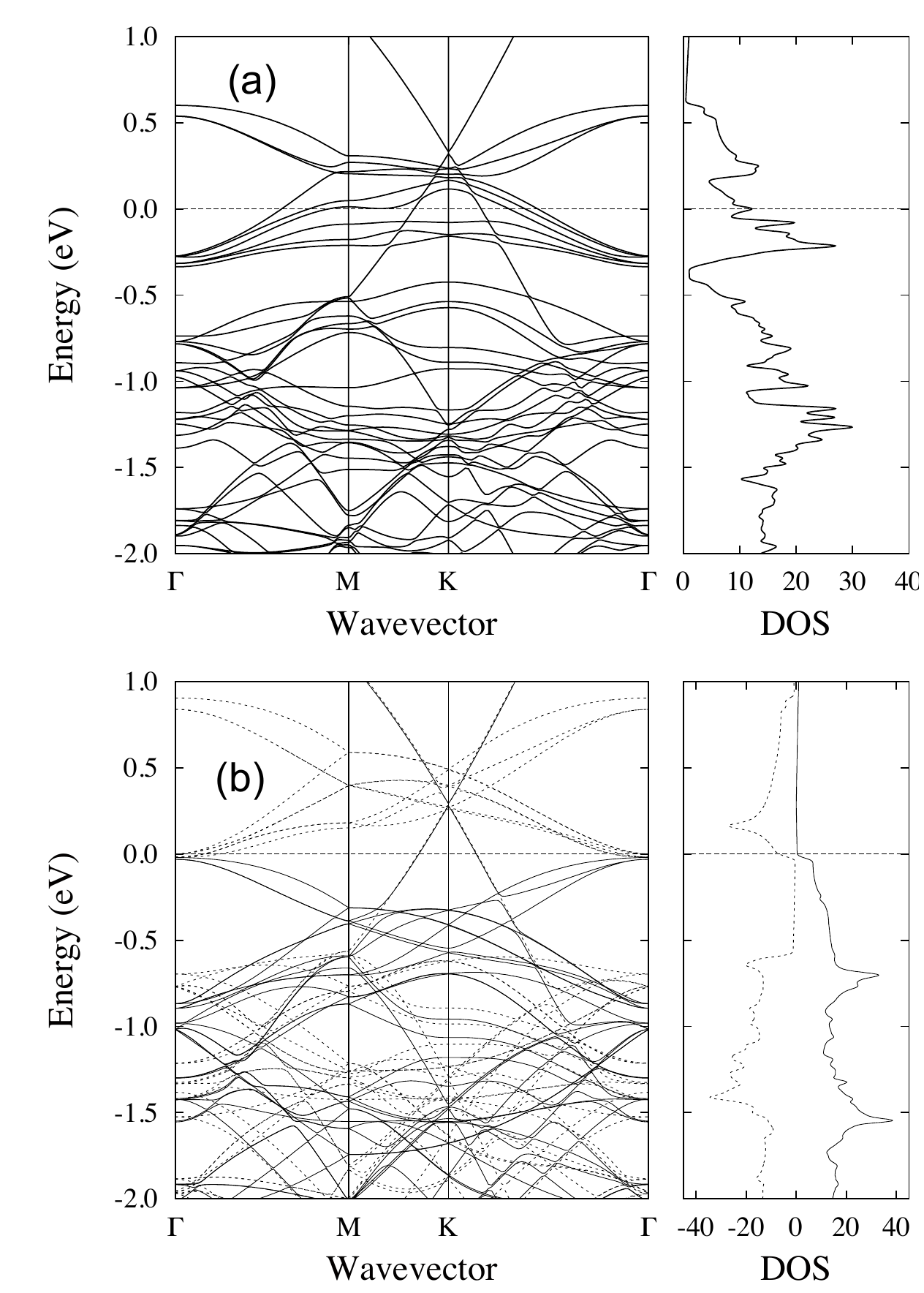}
\caption{Band structure and densities of states of the 3$\times$3 supercell of a NbSe$_{2}$ layer on top of graphene. Panel (a) and (b) correspond to the non-magnetic and magnetic solutions, respectively. Solid and broken lines represent spin-up and spin-down states, respectively.}
\label{Layer-3X3-Graphene}
\end{figure}

\begin{figure}[h]
\includegraphics[width=85mm]
{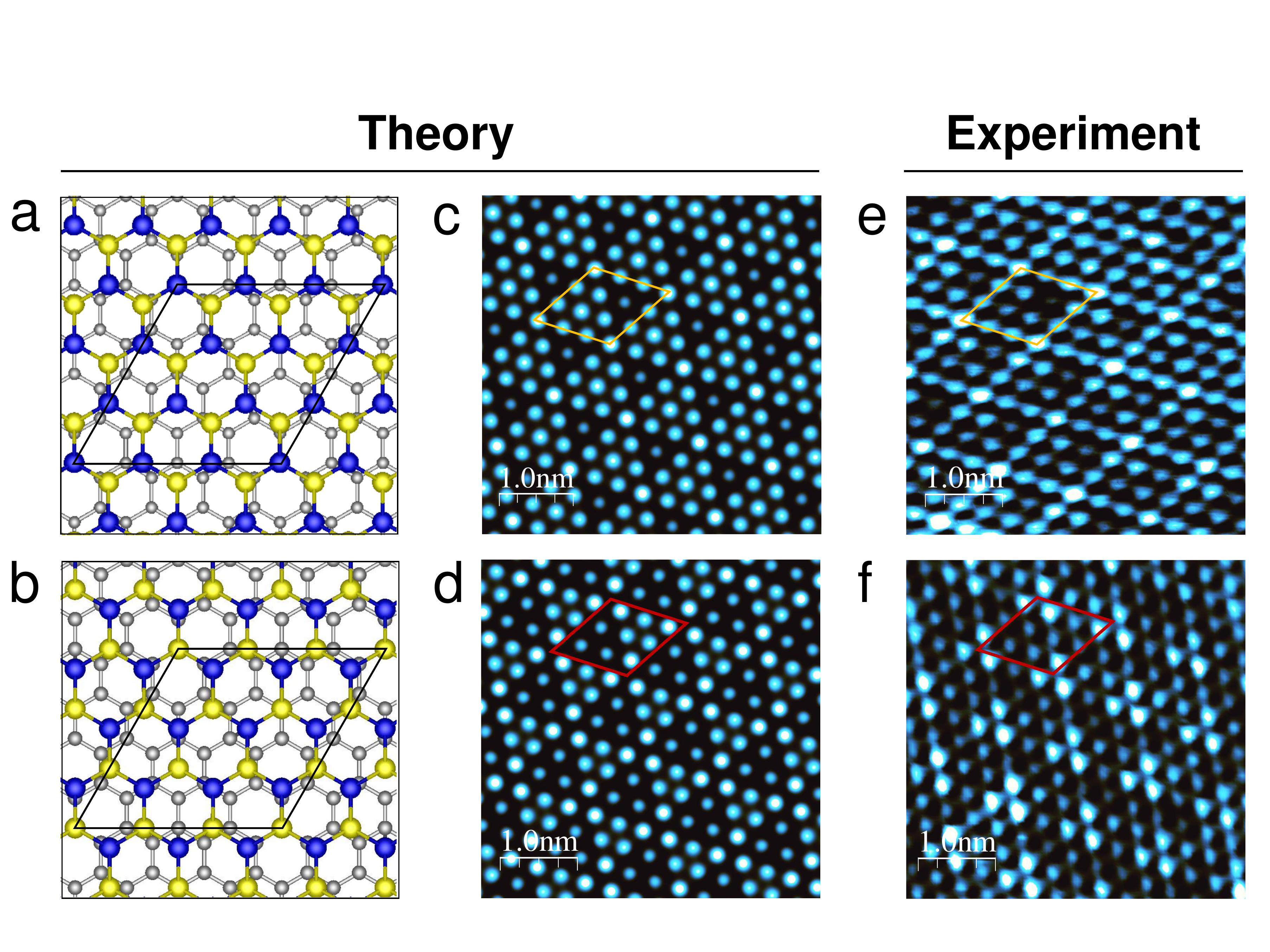}
    \caption{ (Color online) CDW in single-layer NbSe$_{2}$ . Panels (a) and (b) show the atomic position in the equilibrium geometry of the two CDW phases. Grey, yellow and blue stand for C, Nb and Se atoms, respectively. (c) and (d), calculated STM images of the two CDW phases whose atomic structure is shown in panels (a) and (b), respectively. The images are obtained in the Tersoff-Hamann approximation \cite{Tersoff}. The results correspond to empty states in an energy window of 0.01 eV. (e) and (f), experimental constant-current STM experimental images of the two phases taken at T = 340 mK. Parameters: (e) V$_b$ = 85 mV, I$_{t}$ = 370 pA and (e) V$_b$ = 30 mV, I$_{t}$ = 465 pA.} 
    \label{STM}
\end{figure}

\begin{figure}[h]
\includegraphics[width=85mm]
{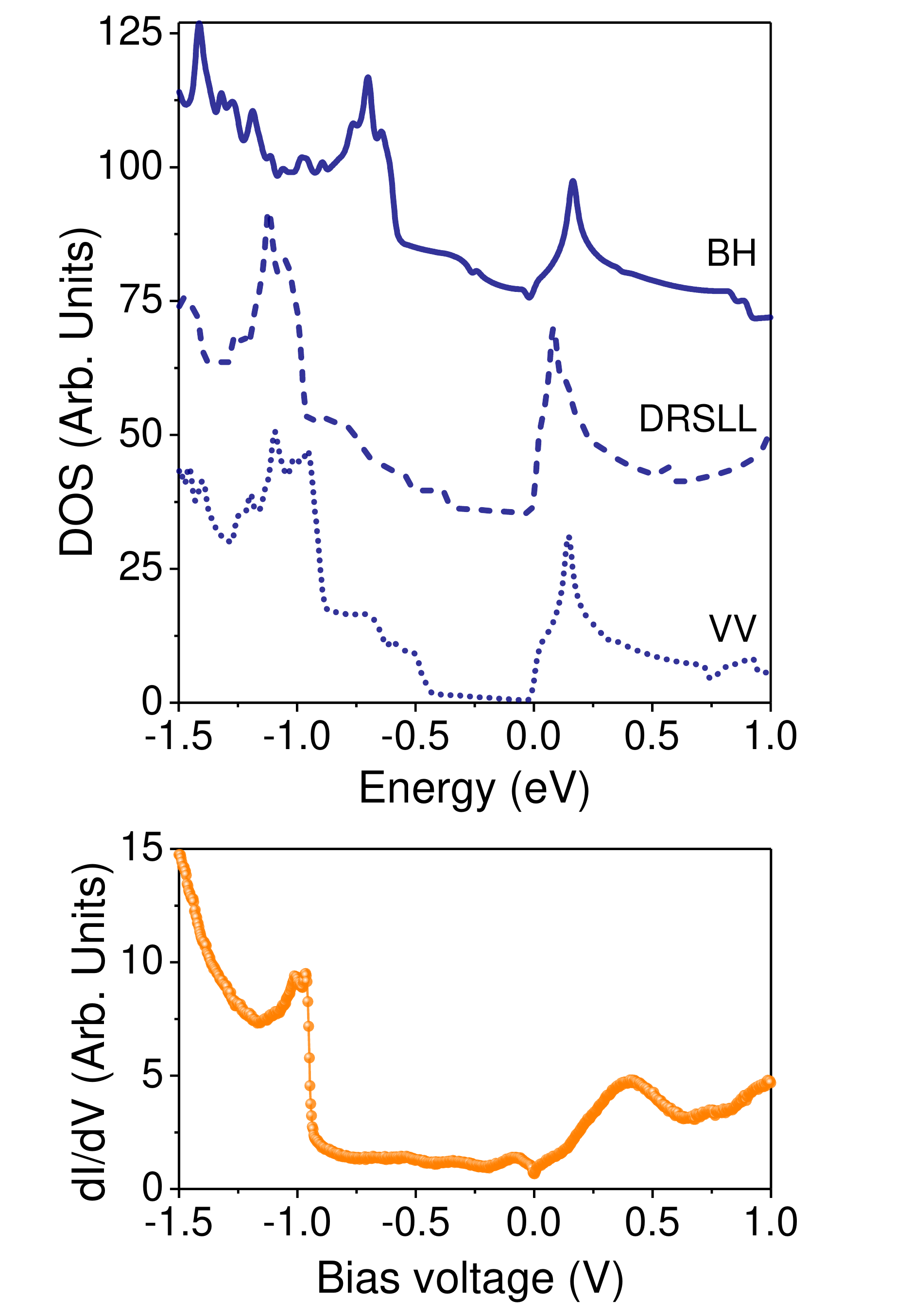}
\caption{(Color on line) Total density of states of a NbSe$_{2}$ layer on top of graphene. The top panel includes the densities of states calculated with three different functionals including van der Waals interactions. BH, VV and DRSLL stand for the functionals of the references, \cite{BH, VV, DRSLL} respectively. The bottom panel shows a typical dI/dV curve (f = 833 Hz, I$_{t}$ = 830 pA, V$_{r.m.s.}$ = 4 mV, T = 4.2 K).}
\label{DOS-vdW}
\end{figure}

For a better simulation of the experimental conditions where the CDW has been visualized in the single layer limit  \cite{Ugeda-NP}, we have introduced a graphene layer below the NbSe$_{2}$ layer, such that there is only a van der Waals interaction between the two materials. For convenience, we have chosen the graphene lattice constant to be commensurate with the lattice constant of NbSe$_{2}$, however as we shall show, this does not perturb the results significantly. The results of the band structure as well as the density of states (DOS) of the 3$\times$3 supercell are shown in Fig. \ref{Layer-3X3-Graphene}. For the non-magnetic calculation, the effect of graphene layer is small. We can identify the linear bands of the graphene cones and also a small charge transfer from the graphene to the NbSe$_{2}$ layers. In addition, we find the calculated DOS and band structure to be similar to those previously calculated \cite{Ordejon-Guinea}. However, the results for the magnetic calculation are different. First, the density of states is similar to the experimental one acquired by scanning tunneling spectroscopy (STS)(see below) and second, due to the asymmetry  induced by the graphene, the top Se layer develops lattice instabilities. The effect of the underneath graphene layer is a transfer of 0.06 electrons to the NbSe$_{2}$  and a breaking of the symmetry between the two Se layers. Depending on the graphene-NbSe$_{2}$ stacking, we find that two different CDW are formed (see Figure \ref{STM}). The energy difference between them is approximately 0.6 meV, with their corresponding densities of states being essentially indistinguishable. As shown in Fig. \ref{STM}, we have experimentally identified these two CDW phases by STM imaging single-layer NbSe$_{2}$ on graphene at low temperature (340 mK). These phases are similar to those previously found in bulk NbSe2 \cite{Ordejon-Ugeda}. The energy gain due to this instability can be estimated to be of the order of 0.4 meV, which is compatible with the CDW critical temperature observed experimentally \cite{Ugeda-NP}. 
Lastly, the total DOS for three different functionals, which include van der Waals interactions, are shown in Fig. \ref{DOS-vdW}, which are compared to our experimental STS curves measured in single-layer NbSe$_{2}$. The theoretical DOS are similar for the three functionals and compare well with the experimental dI/dV spectra measured in single-layer NbSe$_{2}$ on graphene. For occupied states, the theoretical DOS curves show a wide region of low DOS between E$_{F}$ and roughly -0.8 eV followed by a sharp increase due to the presence of several DOS peaks. For unoccupied states, all the theoretical DOS show a broad peak around 0.2 eV. All these features can be identified in the typical dI/dV spectrum shown in figure \ref{DOS-vdW}. The position of the broad peak for unoccupied states is, however, located at 0.4 V. This discrepancy may arise from the poor description of the density functional calculations concerning empty states. This can be improved using LDA+U \cite{LDA+U}, GW \cite{GW}  or similar methods which is beyond our present purposes. We emphasize that all previous DOS calculations for single-layer NbSe$_{2}$ result significantly different from ours as well as from the experimental dI/dV spectra shown here and previously reported for single-layer NbSe$_{2}$ \cite{Ugeda-NP}, which is likely due to lack of spin resolution in previous works compared to the present one.

Experimentally, it is well known that bulk 2H-NbSe$_{2}$ is non-magnetic. However, our first-principles calculations show, for some exchange correlation functionals, that the system is magnetic. In addition, for those functionals that give a correct non-magnetic solution at the equilibrium lattice constant, the system is close to a magnetic solution. However, single-layer NbSe$_{2}$ is magnetic for all the functionals considered in this work and the total magnetic moment is around 1 $\mu_{B}$, which is almost independent of the lattice constant. We note that the possibility of a magnetic solution has been overlooked in the past.

For magnetic solutions we find that the phonon spectrum of single-layer NbSe$_{2}$ does not show soft modes and therefore no CDW, as well as, no atomic distortions. However, a single-layer NbSe$_{2}$ grown on graphene does present two kinds of 3$\times$3 CDW induced by the presence of the underneath graphene layer. Our measurements in single-layer NbSe$_{2}$ confirm the existence of these two CDW phases. 

The presence of magnetism in the 2D limit of NbSe$_{2}$ provide new insight about the intrinsic mechanisms responsible for the experimentally observed superconductivity in NbSe$_{2}$. Further experimental work is therefore needed to confirm the magnetic character of free-standing NbSe$_{2}$ layers and elucidate its interplay with the CDW and superconducting states.

\subsection{Acknowledgements} We  would like to thank Profs. I. Brihuega and S. Vieira for many lively and illuminating discussions. We are indebted to Spanish Ministry of Science and Innovation for financial support through grant FIS2015-64886-C5-5-P and MAT2017-83553-P. M.M.U. acknowledges support by the Spanish MINECO under grant no. MAT2017-88377-C2-1-R and by the ERC Starting grant LINKSPM (Grant 758558).


\begin{thebibliography}{38}
\expandafter\ifx\csname natexlab\endcsname\relax\def\natexlab#1{#1}\fi
\expandafter\ifx\csname bibnamefont\endcsname\relax
  \def\bibnamefont#1{#1}\fi
\expandafter\ifx\csname bibfnamefont\endcsname\relax
  \def\bibfnamefont#1{#1}\fi
\expandafter\ifx\csname citenamefont\endcsname\relax
  \def\citenamefont#1{#1}\fi
\expandafter\ifx\csname url\endcsname\relax
  \def\url#1{\texttt{#1}}\fi
\expandafter\ifx\csname urlprefix\endcsname\relax\def\urlprefix{URL }\fi
\providecommand{\bibinfo}[2]{#2}
\providecommand{\eprint}[2][]{\url{#2}}

\bibitem[{\citenamefont{Wilson et~al.}(1974)\citenamefont{Wilson, Salvo, and
  Mahajan}}]{Wilson-1}
\bibinfo{author}{\bibfnamefont{J.~A.} \bibnamefont{Wilson}},
  \bibinfo{author}{\bibfnamefont{F.~J.~D.} \bibnamefont{Salvo}},
  \bibnamefont{and} \bibinfo{author}{\bibfnamefont{S.}~\bibnamefont{Mahajan}},
  \bibinfo{journal}{Phys. Rev. Lett.} \textbf{\bibinfo{volume}{32}},
  \bibinfo{pages}{882} (\bibinfo{year}{1974}).

\bibitem[{\citenamefont{Wilson et~al.}(1975)\citenamefont{Wilson, Salvo, and
  Mahajan}}]{Wilson-2}
\bibinfo{author}{\bibfnamefont{J.~A.} \bibnamefont{Wilson}},
  \bibinfo{author}{\bibfnamefont{F.~J.~D.} \bibnamefont{Salvo}},
  \bibnamefont{and} \bibinfo{author}{\bibfnamefont{S.}~\bibnamefont{Mahajan}},
  \bibinfo{journal}{Adv. Phys.} \textbf{\bibinfo{volume}{24}},
  \bibinfo{pages}{117} (\bibinfo{year}{1975}).

\bibitem[{\citenamefont{Weber et~al.}(2013)\citenamefont{Weber, Hott, Heid,
  Bohnen, Rosenkranz, Castellan, Osborn, Said, Leu, and Reznik}}]{Weber-PRB}
\bibinfo{author}{\bibfnamefont{F.}~\bibnamefont{Weber}},
  \bibinfo{author}{\bibfnamefont{R.}~\bibnamefont{Hott}},
  \bibinfo{author}{\bibfnamefont{R.}~\bibnamefont{Heid}},
  \bibinfo{author}{\bibfnamefont{K.-P.} \bibnamefont{Bohnen}},
  \bibinfo{author}{\bibfnamefont{S.}~\bibnamefont{Rosenkranz}},
  \bibinfo{author}{\bibfnamefont{J.-P.} \bibnamefont{Castellan}},
  \bibinfo{author}{\bibfnamefont{R.}~\bibnamefont{Osborn}},
  \bibinfo{author}{\bibfnamefont{A.~H.} \bibnamefont{Said}},
  \bibinfo{author}{\bibfnamefont{B.~M.} \bibnamefont{Leu}}, \bibnamefont{and}
  \bibinfo{author}{\bibfnamefont{D.}~\bibnamefont{Reznik}},
  \bibinfo{journal}{Phys. Rev. B} \textbf{\bibinfo{volume}{87}},
  \bibinfo{pages}{245111} (\bibinfo{year}{2013}).

\bibitem[{\citenamefont{Weber et~al.}(2011)\citenamefont{Weber, Rosenkranz,
  Castellan, Osborn, Hott, Heid, Bohnen, Egami, Said, and Reznik}}]{Weber-PRL}
\bibinfo{author}{\bibfnamefont{F.}~\bibnamefont{Weber}},
  \bibinfo{author}{\bibfnamefont{S.}~\bibnamefont{Rosenkranz}},
  \bibinfo{author}{\bibfnamefont{J.-P.} \bibnamefont{Castellan}},
  \bibinfo{author}{\bibfnamefont{R.}~\bibnamefont{Osborn}},
  \bibinfo{author}{\bibfnamefont{R.}~\bibnamefont{Hott}},
  \bibinfo{author}{\bibfnamefont{R.}~\bibnamefont{Heid}},
  \bibinfo{author}{\bibfnamefont{K.-P.} \bibnamefont{Bohnen}},
  \bibinfo{author}{\bibfnamefont{T.}~\bibnamefont{Egami}},
  \bibinfo{author}{\bibfnamefont{A.~H.} \bibnamefont{Said}}, \bibnamefont{and}
  \bibinfo{author}{\bibfnamefont{D.}~\bibnamefont{Reznik}},
  \bibinfo{journal}{Phys. Rev. Lett.} \textbf{\bibinfo{volume}{107}},
  \bibinfo{pages}{107403} (\bibinfo{year}{2011}).

\bibitem[{\citenamefont{Johannes et~al.}(2006)\citenamefont{Johannes, Mazin,
  and Howells}}]{Mazin-1}
\bibinfo{author}{\bibfnamefont{M.~D.} \bibnamefont{Johannes}},
  \bibinfo{author}{\bibfnamefont{I.~I.} \bibnamefont{Mazin}}, \bibnamefont{and}
  \bibinfo{author}{\bibfnamefont{C.~A.} \bibnamefont{Howells}},
  \bibinfo{journal}{Phys. Rev. B} \textbf{\bibinfo{volume}{73}},
  \bibinfo{pages}{205102} (\bibinfo{year}{2006}).

\bibitem[{\citenamefont{Johannes and Mazin}(2008)}]{Mazin-2}
\bibinfo{author}{\bibfnamefont{M.~D.} \bibnamefont{Johannes}} \bibnamefont{and}
  \bibinfo{author}{\bibfnamefont{I.~I.} \bibnamefont{Mazin}},
  \bibinfo{journal}{Phys. Rev. B} \textbf{\bibinfo{volume}{77}},
  \bibinfo{pages}{16135} (\bibinfo{year}{2008}).

\bibitem[{\citenamefont{Ugeda et~al.}(2016)\citenamefont{Ugeda, Bradley, Zhang,
  Onishi, Chen, Ruan, Ojeda-Aristizibal, Ryu, Edmonds, Tsai et~al.}}]{Ugeda-NP}
\bibinfo{author}{\bibfnamefont{M.~M.} \bibnamefont{Ugeda}},
  \bibinfo{author}{\bibfnamefont{A.~J.} \bibnamefont{Bradley}},
  \bibinfo{author}{\bibfnamefont{Y.}~\bibnamefont{Zhang}},
  \bibinfo{author}{\bibfnamefont{S.}~\bibnamefont{Onishi}},
  \bibinfo{author}{\bibfnamefont{Y.}~\bibnamefont{Chen}},
  \bibinfo{author}{\bibfnamefont{W.}~\bibnamefont{Ruan}},
  \bibinfo{author}{\bibfnamefont{C.}~\bibnamefont{Ojeda-Aristizibal}},
  \bibinfo{author}{\bibfnamefont{H.}~\bibnamefont{Ryu}},
  \bibinfo{author}{\bibfnamefont{M.~T.} \bibnamefont{Edmonds}},
  \bibinfo{author}{\bibfnamefont{H.-Z.} \bibnamefont{Tsai}},
  \bibnamefont{et~al.}, \bibinfo{journal}{Nature Physics}
  \textbf{\bibinfo{volume}{12}}, \bibinfo{pages}{92} (\bibinfo{year}{2016}).

\bibitem[{\citenamefont{Calandra et~al.}(2009)\citenamefont{Calandra, Mazin,
  and Mauri}}]{Calandra-few-layer}
\bibinfo{author}{\bibfnamefont{M.}~\bibnamefont{Calandra}},
  \bibinfo{author}{\bibfnamefont{I.~I.} \bibnamefont{Mazin}}, \bibnamefont{and}
  \bibinfo{author}{\bibfnamefont{F.}~\bibnamefont{Mauri}},
  \bibinfo{journal}{Phys. Rev. B} \textbf{\bibinfo{volume}{80}},
  \bibinfo{pages}{241108R} (\bibinfo{year}{2009}).

\bibitem[{\citenamefont{Guster et~al.}(2019)\citenamefont{Guster,
  Rubio-Verd\'{u}, Robles, Zald\'{i}var, Dreher, Pruneda, Silva-Guill\'{e}n,
  Choi, Pascual, Ugeda et~al.}}]{Ordejon-Ugeda}
\bibinfo{author}{\bibfnamefont{B.}~\bibnamefont{Guster}},
  \bibinfo{author}{\bibfnamefont{C.}~\bibnamefont{Rubio-Verd\'{u}}},
  \bibinfo{author}{\bibfnamefont{R.}~\bibnamefont{Robles}},
  \bibinfo{author}{\bibfnamefont{J.}~\bibnamefont{Zald\'{i}var}},
  \bibinfo{author}{\bibfnamefont{P.}~\bibnamefont{Dreher}},
  \bibinfo{author}{\bibfnamefont{M.}~\bibnamefont{Pruneda}},
  \bibinfo{author}{\bibfnamefont{J.~A.} \bibnamefont{Silva-Guill\'{e}n}},
  \bibinfo{author}{\bibfnamefont{D.-J.} \bibnamefont{Choi}},
  \bibinfo{author}{\bibfnamefont{J.~I.} \bibnamefont{Pascual}},
  \bibinfo{author}{\bibfnamefont{M.~M.} \bibnamefont{Ugeda}},
  \bibnamefont{et~al.}, \bibinfo{journal}{Nano Letters}
  \textbf{\bibinfo{volume}{19}}, \bibinfo{pages}{3027} (\bibinfo{year}{2019}).

\bibitem[{\citenamefont{Zhou et~al.}(2013)\citenamefont{Zhou, Yang, Xiang, and
  Zu}}]{Magnetism-TM-Se2}
\bibinfo{author}{\bibfnamefont{Y.}~\bibnamefont{Zhou}},
  \bibinfo{author}{\bibfnamefont{C.}~\bibnamefont{Yang}},
  \bibinfo{author}{\bibfnamefont{X.}~\bibnamefont{Xiang}}, \bibnamefont{and}
  \bibinfo{author}{\bibfnamefont{X.}~\bibnamefont{Zu}}, \bibinfo{journal}{Phys.
  Chem. Chem. Phys.} \textbf{\bibinfo{volume}{15}}, \bibinfo{pages}{14202}
  (\bibinfo{year}{2013}).

\bibitem[{\citenamefont{Ma et~al.}(2012)\citenamefont{Ma, Dai, Guo, Niu, Zhu,
  and Huang}}]{Ma}
\bibinfo{author}{\bibfnamefont{Y.}~\bibnamefont{Ma}},
  \bibinfo{author}{\bibfnamefont{Y.}~\bibnamefont{Dai}},
  \bibinfo{author}{\bibfnamefont{M.}~\bibnamefont{Guo}},
  \bibinfo{author}{\bibfnamefont{C.}~\bibnamefont{Niu}},
  \bibinfo{author}{\bibfnamefont{Y.}~\bibnamefont{Zhu}}, \bibnamefont{and}
  \bibinfo{author}{\bibfnamefont{B.}~\bibnamefont{Huang}},
  \bibinfo{journal}{ACS Nano} \textbf{\bibinfo{volume}{6}},
  \bibinfo{pages}{1695} (\bibinfo{year}{2012}).

\bibitem[{\citenamefont{Ataca et~al.}(2012)\citenamefont{Ataca, H.Sahin, and
  Ciraci}}]{Attaca}
\bibinfo{author}{\bibfnamefont{C.}~\bibnamefont{Ataca}},
  \bibinfo{author}{\bibnamefont{H.Sahin}}, \bibnamefont{and}
  \bibinfo{author}{\bibfnamefont{S.}~\bibnamefont{Ciraci}},
  \bibinfo{journal}{Journal of Physical Chemistry C}
  \textbf{\bibinfo{volume}{116}}, \bibinfo{pages}{8983} (\bibinfo{year}{2012}).

\bibitem[{\citenamefont{M.Bonilla et~al.}(2018)\citenamefont{M.Bonilla,
  Kolekar, Ma, Diaz, Kalappattil, Das, Eggers, Gutierrez, Phan, and
  Batzill}}]{Batzill}
\bibinfo{author}{\bibnamefont{M.Bonilla}},
  \bibinfo{author}{\bibfnamefont{S.}~\bibnamefont{Kolekar}},
  \bibinfo{author}{\bibfnamefont{Y.}~\bibnamefont{Ma}},
  \bibinfo{author}{\bibfnamefont{H.~C.} \bibnamefont{Diaz}},
  \bibinfo{author}{\bibfnamefont{V.}~\bibnamefont{Kalappattil}},
  \bibinfo{author}{\bibfnamefont{R.}~\bibnamefont{Das}},
  \bibinfo{author}{\bibfnamefont{T.}~\bibnamefont{Eggers}},
  \bibinfo{author}{\bibfnamefont{H.~R.} \bibnamefont{Gutierrez}},
  \bibinfo{author}{\bibfnamefont{M.-H.} \bibnamefont{Phan}}, \bibnamefont{and}
  \bibinfo{author}{\bibfnamefont{M.}~\bibnamefont{Batzill}},
  \bibinfo{journal}{Nature Nanotechnology} \textbf{\bibinfo{volume}{13}},
  \bibinfo{pages}{289} (\bibinfo{year}{2018}).

\bibitem[{\citenamefont{Fumega et~al.}(2019)\citenamefont{Fumega, Gobbi,
  Dreher, Wan, Gonz\'{a}lez-Orellana, Pena-D\'{i}?az, Rogero, Herrero-Mart??n,
  Gargiani, Ilyn et~al.}}]{Fumega}
\bibinfo{author}{\bibfnamefont{A.~O.} \bibnamefont{Fumega}},
  \bibinfo{author}{\bibfnamefont{M.}~\bibnamefont{Gobbi}},
  \bibinfo{author}{\bibfnamefont{P.}~\bibnamefont{Dreher}},
  \bibinfo{author}{\bibfnamefont{W.}~\bibnamefont{Wan}},
  \bibinfo{author}{\bibfnamefont{C.}~\bibnamefont{Gonz\'{a}lez-Orellana}},
  \bibinfo{author}{\bibfnamefont{M.}~\bibnamefont{Pena-D\'{i}?az}},
  \bibinfo{author}{\bibfnamefont{C.}~\bibnamefont{Rogero}},
  \bibinfo{author}{\bibfnamefont{J.}~\bibnamefont{Herrero-Mart??n}},
  \bibinfo{author}{\bibfnamefont{P.}~\bibnamefont{Gargiani}},
  \bibinfo{author}{\bibfnamefont{M.}~\bibnamefont{Ilyn}}, \bibnamefont{et~al.},
  \bibinfo{journal}{Journal of Physical Chemistry C}
  \textbf{\bibinfo{volume}{123}}, \bibinfo{pages}{27802}
  (\bibinfo{year}{2019}).

\bibitem[{\citenamefont{O'ÄôHara et~al.}(2018)\citenamefont{O'ÄôHara, Zhu,
  Trout, Ahmed, Luo, Lee, Brenner, Rajan, Gupta, McComb et~al.}}]{Ohara}
\bibinfo{author}{\bibfnamefont{D.~J.} \bibnamefont{O'ÄôHara}},
  \bibinfo{author}{\bibfnamefont{T.}~\bibnamefont{Zhu}},
  \bibinfo{author}{\bibfnamefont{A.~H.} \bibnamefont{Trout}},
  \bibinfo{author}{\bibfnamefont{A.~S.} \bibnamefont{Ahmed}},
  \bibinfo{author}{\bibfnamefont{Y.~K.} \bibnamefont{Luo}},
  \bibinfo{author}{\bibfnamefont{C.~H.} \bibnamefont{Lee}},
  \bibinfo{author}{\bibfnamefont{M.~R.} \bibnamefont{Brenner}},
  \bibinfo{author}{\bibfnamefont{S.}~\bibnamefont{Rajan}},
  \bibinfo{author}{\bibfnamefont{J.~A.} \bibnamefont{Gupta}},
  \bibinfo{author}{\bibfnamefont{D.~W.} \bibnamefont{McComb}},
  \bibnamefont{et~al.}, \bibinfo{journal}{Nano Letters}
  \textbf{\bibinfo{volume}{18}}, \bibinfo{pages}{3125} (\bibinfo{year}{2018}).

\bibitem[{\citenamefont{Avsar et~al.}(2019)\citenamefont{Avsar, Ciarrocchi,
  Pizzochero, Unuchek, Yazyev, and Kis}}]{Avsar}
\bibinfo{author}{\bibfnamefont{A.}~\bibnamefont{Avsar}},
  \bibinfo{author}{\bibfnamefont{A.}~\bibnamefont{Ciarrocchi}},
  \bibinfo{author}{\bibfnamefont{M.}~\bibnamefont{Pizzochero}},
  \bibinfo{author}{\bibfnamefont{D.}~\bibnamefont{Unuchek}},
  \bibinfo{author}{\bibfnamefont{O.~V.} \bibnamefont{Yazyev}},
  \bibnamefont{and} \bibinfo{author}{\bibfnamefont{A.}~\bibnamefont{Kis}},
  \bibinfo{journal}{Nature Nanotechnology} \textbf{\bibinfo{volume}{14}},
  \bibinfo{pages}{674} (\bibinfo{year}{2019}).

\bibitem[{\citenamefont{Zhou et~al.}(2012)\citenamefont{Zhou, Wang, Yang, Zu,
  Sun, and Gao}}]{Strain}
\bibinfo{author}{\bibfnamefont{Y.}~\bibnamefont{Zhou}},
  \bibinfo{author}{\bibfnamefont{Z.}~\bibnamefont{Wang}},
  \bibinfo{author}{\bibfnamefont{P.}~\bibnamefont{Yang}},
  \bibinfo{author}{\bibfnamefont{X.}~\bibnamefont{Zu}},
  \bibinfo{author}{\bibfnamefont{X.}~\bibnamefont{Sun}}, \bibnamefont{and}
  \bibinfo{author}{\bibfnamefont{F.}~\bibnamefont{Gao}}, \bibinfo{journal}{ACS
  Nano} \textbf{\bibinfo{volume}{6}}, \bibinfo{pages}{9736}
  (\bibinfo{year}{2012}).

\bibitem[{\citenamefont{Zheng et~al.}(2018)\citenamefont{Zheng, Zhou, Liu, and
  Feng}}]{Zheng}
\bibinfo{author}{\bibfnamefont{F.}~\bibnamefont{Zheng}},
  \bibinfo{author}{\bibfnamefont{Z.}~\bibnamefont{Zhou}},
  \bibinfo{author}{\bibfnamefont{X.}~\bibnamefont{Liu}}, \bibnamefont{and}
  \bibinfo{author}{\bibfnamefont{J.}~\bibnamefont{Feng}},
  \bibinfo{journal}{Phys. Rev. B} \textbf{\bibinfo{volume}{97}},
  \bibinfo{pages}{08101} (\bibinfo{year}{2018}).

\bibitem[{\citenamefont{Hohenberg and Kohn}(1964)}]{DFT1}
\bibinfo{author}{\bibfnamefont{P.}~\bibnamefont{Hohenberg}} \bibnamefont{and}
  \bibinfo{author}{\bibfnamefont{W.}~\bibnamefont{Kohn}},
  \bibinfo{journal}{Phys. Rev.} \textbf{\bibinfo{volume}{136}},
  \bibinfo{pages}{B864} (\bibinfo{year}{1964}).

\bibitem[{\citenamefont{Kohn and Sham}(1965)}]{DFT2}
\bibinfo{author}{\bibfnamefont{W.}~\bibnamefont{Kohn}} \bibnamefont{and}
  \bibinfo{author}{\bibfnamefont{L.~J.} \bibnamefont{Sham}},
  \bibinfo{journal}{Phys. Rev.} \textbf{\bibinfo{volume}{140}},
  \bibinfo{pages}{A1133} (\bibinfo{year}{1965}).

\bibitem[{\citenamefont{Ordej\'{o}n et~al.}(1996)\citenamefont{Ordej\'{o}n,
  Artacho, and Soler}}]{Siesta1}
\bibinfo{author}{\bibfnamefont{P.}~\bibnamefont{Ordej\'{o}n}},
  \bibinfo{author}{\bibfnamefont{E.}~\bibnamefont{Artacho}}, \bibnamefont{and}
  \bibinfo{author}{\bibfnamefont{J.~M.} \bibnamefont{Soler}},
  \bibinfo{journal}{Phys. Rev. B} \textbf{\bibinfo{volume}{53}},
  \bibinfo{pages}{R10441} (\bibinfo{year}{1996}).

\bibitem[{\citenamefont{Soler et~al.}(2002)\citenamefont{Soler, Artacho, Gale,
  Garc\'{i}a, Junquera, Ordej\'{o}n, and Sanchez-Portal}}]{Siesta2}
\bibinfo{author}{\bibfnamefont{J.}~\bibnamefont{Soler}},
  \bibinfo{author}{\bibfnamefont{E.}~\bibnamefont{Artacho}},
  \bibinfo{author}{\bibfnamefont{J.}~\bibnamefont{Gale}},
  \bibinfo{author}{\bibfnamefont{A.}~\bibnamefont{Garc\'{i}a}},
  \bibinfo{author}{\bibfnamefont{J.}~\bibnamefont{Junquera}},
  \bibinfo{author}{\bibfnamefont{P.}~\bibnamefont{Ordej\'{o}n}},
  \bibnamefont{and}
  \bibinfo{author}{\bibfnamefont{D.}~\bibnamefont{Sanchez-Portal}},
  \bibinfo{journal}{J. Phys.: Condens. Matter} \textbf{\bibinfo{volume}{14}},
  \bibinfo{pages}{2745} (\bibinfo{year}{2002}).

\bibitem[{\citenamefont{Sankey and Niklewski}(1989)}]{Orbitals}
\bibinfo{author}{\bibfnamefont{O.~F.} \bibnamefont{Sankey}} \bibnamefont{and}
  \bibinfo{author}{\bibfnamefont{D.~J.} \bibnamefont{Niklewski}},
  \bibinfo{journal}{Phys. Rev. B} \textbf{\bibinfo{volume}{40}},
  \bibinfo{pages}{3979} (\bibinfo{year}{1989}).

\bibitem[{\citenamefont{Perdew and Wang}(1992)}]{GGA}
\bibinfo{author}{\bibfnamefont{J.~P.} \bibnamefont{Perdew}} \bibnamefont{and}
  \bibinfo{author}{\bibfnamefont{Y.}~\bibnamefont{Wang}},
  \bibinfo{journal}{Phys. Rev. B} \textbf{\bibinfo{volume}{45}},
  \bibinfo{pages}{13244} (\bibinfo{year}{1992}).

\bibitem[{\citenamefont{Berland and Hyldgaard}(2014)}]{BH}
\bibinfo{author}{\bibfnamefont{K.}~\bibnamefont{Berland}} \bibnamefont{and}
  \bibinfo{author}{\bibfnamefont{P.}~\bibnamefont{Hyldgaard}},
  \bibinfo{journal}{Phys. Rev. B} \textbf{\bibinfo{volume}{89}},
  \bibinfo{pages}{035412} (\bibinfo{year}{2014}).

\bibitem[{\citenamefont{Rom\'{a}n-P\'{e}rez and Soler}(2010)}]{Roman}
\bibinfo{author}{\bibfnamefont{G.}~\bibnamefont{Rom\'{a}n-P\'{e}rez}}
  \bibnamefont{and} \bibinfo{author}{\bibfnamefont{J.~M.} \bibnamefont{Soler}},
  \bibinfo{journal}{Phys. Rev. Lett.} \textbf{\bibinfo{volume}{103}},
  \bibinfo{pages}{096102} (\bibinfo{year}{2010}).

\bibitem[{\citenamefont{Methfessel and Paxton}(1989)}]{smearing}
\bibinfo{author}{\bibfnamefont{M.}~\bibnamefont{Methfessel}} \bibnamefont{and}
  \bibinfo{author}{\bibfnamefont{A.~T.} \bibnamefont{Paxton}},
  \bibinfo{journal}{Phys. Rev. B} \textbf{\bibinfo{volume}{40}},
  \bibinfo{pages}{3616} (\bibinfo{year}{1989}).

\bibitem[{SM()}]{SM}
\bibinfo{note}{See Suplemental Material}.

\bibitem[{\citenamefont{Perdew and Zunger}(1981)}]{LDA2}
\bibinfo{author}{\bibfnamefont{J.~P.} \bibnamefont{Perdew}} \bibnamefont{and}
  \bibinfo{author}{\bibfnamefont{A.}~\bibnamefont{Zunger}},
  \bibinfo{journal}{Phys. Rev. B} \textbf{\bibinfo{volume}{23}},
  \bibinfo{pages}{5048} (\bibinfo{year}{1981}).

\bibitem[{\citenamefont{Meerschaut and Deudon}(2001)}]{Meerschaut}
\bibinfo{author}{\bibfnamefont{A.}~\bibnamefont{Meerschaut}} \bibnamefont{and}
  \bibinfo{author}{\bibfnamefont{C.}~\bibnamefont{Deudon}},
  \bibinfo{journal}{Materials Research Bulletin} \textbf{\bibinfo{volume}{36}},
  \bibinfo{pages}{1721} (\bibinfo{year}{2001}).

\bibitem[{\citenamefont{Kim and Son}(2017)}]{Spin-Orbit}
\bibinfo{author}{\bibfnamefont{S.}~\bibnamefont{Kim}} \bibnamefont{and}
  \bibinfo{author}{\bibfnamefont{Y.-W.} \bibnamefont{Son}},
  \bibinfo{journal}{Phys. Rev. B} \textbf{\bibinfo{volume}{96}},
  \bibinfo{pages}{155439} (\bibinfo{year}{2017}).

\bibitem[{\citenamefont{Silva-Guill\'{e}n
  et~al.}(2016)\citenamefont{Silva-Guill\'{e}n, Ordej\'{o}n, Guinea, and
  Canadell}}]{Ordejon-Guinea}
\bibinfo{author}{\bibfnamefont{J.~A.} \bibnamefont{Silva-Guill\'{e}n}},
  \bibinfo{author}{\bibfnamefont{P.}~\bibnamefont{Ordej\'{o}n}},
  \bibinfo{author}{\bibfnamefont{F.}~\bibnamefont{Guinea}}, \bibnamefont{and}
  \bibinfo{author}{\bibfnamefont{E.}~\bibnamefont{Canadell}},
  \bibinfo{journal}{2D Materials} \textbf{\bibinfo{volume}{3}},
  \bibinfo{pages}{035028} (\bibinfo{year}{2016}).

\bibitem[{\citenamefont{Bianco et~al.}(2020)\citenamefont{Bianco, Monacelli,
  Calandra, Mauri, and Errea}}]{ion}
\bibinfo{author}{\bibfnamefont{R.}~\bibnamefont{Bianco}},
  \bibinfo{author}{\bibfnamefont{L.}~\bibnamefont{Monacelli}},
  \bibinfo{author}{\bibfnamefont{M.}~\bibnamefont{Calandra}},
  \bibinfo{author}{\bibfnamefont{F.}~\bibnamefont{Mauri}}, \bibnamefont{and}
  \bibinfo{author}{\bibfnamefont{I.}~\bibnamefont{Errea}},
  \bibinfo{journal}{arXiv:2004.08147} \textbf{\bibinfo{volume}{xx}},
  \bibinfo{pages}{xx} (\bibinfo{year}{2020}).

\bibitem[{\citenamefont{Tersoff and Hamann}(1983)}]{Tersoff}
\bibinfo{author}{\bibfnamefont{J.}~\bibnamefont{Tersoff}} \bibnamefont{and}
  \bibinfo{author}{\bibfnamefont{D.~R.} \bibnamefont{Hamann}},
  \bibinfo{journal}{Phys. Rev. Lett.} \textbf{\bibinfo{volume}{50}},
  \bibinfo{pages}{1998} (\bibinfo{year}{1983}).

\bibitem[{\citenamefont{Himmetoglu et~al.}(2014)\citenamefont{Himmetoglu,
  Floris, de~Gironcoli, and Cococcioni}}]{LDA+U}
\bibinfo{author}{\bibfnamefont{B.}~\bibnamefont{Himmetoglu}},
  \bibinfo{author}{\bibfnamefont{A.}~\bibnamefont{Floris}},
  \bibinfo{author}{\bibfnamefont{S.}~\bibnamefont{de~Gironcoli}},
  \bibnamefont{and}
  \bibinfo{author}{\bibfnamefont{M.}~\bibnamefont{Cococcioni}},
  \bibinfo{journal}{International Journal of Quantum Chemistry}
  \textbf{\bibinfo{volume}{114}}, \bibinfo{pages}{14} (\bibinfo{year}{2014}).

\bibitem[{\citenamefont{Hybertsen and Louie}(1986)}]{GW}
\bibinfo{author}{\bibfnamefont{M.~S.} \bibnamefont{Hybertsen}}
  \bibnamefont{and} \bibinfo{author}{\bibfnamefont{S.~G.} \bibnamefont{Louie}},
  \bibinfo{journal}{Phys. Rev. B} \textbf{\bibinfo{volume}{34}},
  \bibinfo{pages}{5390} (\bibinfo{year}{1986}).

\bibitem[{\citenamefont{Vydrov and Voorhis}(2010)}]{VV}
\bibinfo{author}{\bibfnamefont{O.~A.} \bibnamefont{Vydrov}} \bibnamefont{and}
  \bibinfo{author}{\bibfnamefont{T.~V.} \bibnamefont{Voorhis}},
  \bibinfo{journal}{J. Chem. Phys.} \textbf{\bibinfo{volume}{133}},
  \bibinfo{pages}{244103} (\bibinfo{year}{2010}).

\bibitem[{\citenamefont{Dion et~al.}(2004)\citenamefont{Dion, Rydberg,
  Schroder, Langreth, and Lundqvist}}]{DRSLL}
\bibinfo{author}{\bibfnamefont{M.}~\bibnamefont{Dion}},
  \bibinfo{author}{\bibfnamefont{H.}~\bibnamefont{Rydberg}},
  \bibinfo{author}{\bibfnamefont{E.}~\bibnamefont{Schroder}},
  \bibinfo{author}{\bibfnamefont{D.~C.} \bibnamefont{Langreth}},
  \bibnamefont{and} \bibinfo{author}{\bibfnamefont{B.~I.}
  \bibnamefont{Lundqvist}}, \bibinfo{journal}{Phys. Rev. Lett.}
  \textbf{\bibinfo{volume}{103}}, \bibinfo{pages}{096102}
  (\bibinfo{year}{2004}).

\end{thebibliography}

\end{document}